# Quasi-monochromatic bright gamma-ray generation from synchronized Compton scattering via azimuthal spatial-temporal coupling

Xuesong Geng[1], Liangliang Ji[1,*], Baifei Shen[1,2,†]

[1]*State Key Laboratory of High Field Laser Physics and CAS Center for Excellence in Ultra-intense Laser Science, Shanghai Institute of Optics and Fine Mechanics (SIOM), Chinese Academy of Sciences (CAS), Shanghai 201800, China.*
[2]*Shanghai Normal University, Shanghai 200234, China.*

*jill@siom.ac.cn

†bfshen@mail.shcnc.ac.cn

High energy photons can be generated via inverse Compton scattering (ICS) in the collision between energetic electrons and intense laser pulse. The development of laser plasma accelerators promises compact and all-optical gamma-ray sources by colliding the electrons from laser wakefield accelerators to its high-power driving pulse reflected by a plasma mirror. However, the law of optical focusing hinders realization of both high photon yield and monochromatic spectrum in this scenario. We propose an azimuthal spatial-temporal convertor that decouples the focal field strength from laser spot size using helical parabolic geometry. It decomposes the driving laser beam into a pulse train of almost identical divergence angle and focal depth, creating synchronized ICS in the optimized linear regime. The scheme resolves the dilemma between high efficiency and narrow energy spread, facilitating the generation of monochromatic gamma-ray using high power lasers beyond relativistic field strengths.

## I. INTRODUCTION

Photons in the gamma-ray regime can stimulate photonuclear reactions, providing a powerful tool to probe the dynamics in nuclei. To this end, it is often required that the gamma-ray beam is mono-chromatic with sufficient photon flux to selectively excite the nuclear states [1]. This has been a major quest in developing nuclear photonics but faces significant challenges. Synchrotron radiation [2–6] and free-electron laser [7–9] have been quite successful in providing bright light sources up to hard-x-ray regime. However, it is unlikely to extend the state-of-the-art techniques to the gamma-ray regime. Instead, abundant gamma-photons can be efficiently produced via bremsstrahlung [10], where the photon spectra usually follow exponential decay distribution, and the beam is rather diverged [11]. On the other hand, the Inverse Compton Scattering (ICS) [12,13] mechanism promises highly directional gamma-beams with narrow energy spread, via low energy laser photons scattered off relativistic electrons. They gain a maximum of $4\gamma^2$ ($\gamma$ the electron Lorentz factor) boost to the laser frequency and become blue-shifted [14,15].

Various ICS gamma-ray sources have been realized worldwide in conventional accelerators [16–19] or via laser-driven wakefield accelerated (LWFA) electrons [20–

22]. The latter is particularly promising since it generates ultra-bright ultra-fast gamma-ray flashes in a compact all-optical scenario. Laser-based ICS is usually facilitated via the reflection-collision geometry: the high-power femtosecond laser pulse driving wakefield acceleration also serves as the collision laser after reflected by a plasma mirror. This approach takes full advantage of the ultra-high photon density provided by the driving laser and resolves the synchronization issue between the laser and electron beams [22–26].

Typical photon number generated in ICS is about $10^{7-9}$ per pulse [27,22,23], which is much smaller than the one from bremsstrahlung (~$10^{10-11}$). One may increase the light intensity of the scattering laser in ICS to boost the photon yield. However, this is limited by the nonlinear threshold. When the field strength of the collision laser exceeds $a_0 = 1$, ICS enters the nonlinear or the multi-photon absorption regime, where the gamma ray spectrum is red-shifted and broadened [28–30]. Here $a_0 = E_0 e/\omega m c$ denotes the nonlinearity of the collision process, with $E_0$ the laser electric magnitude, $e$ the unit charge, $\omega$ the laser frequency, $m$ the electron mass and $c$ the speed of light. In other words, there exists a dilemma between photon yield and spectrum width. Take a typical 50 TW / 30 fs laser pulse in LWFA as an example. The linear ICS ($a_0 < 1$) requires the beam waist larger than $w_0 \approx 40\mu m$ as shown in red in Fig. 1(a). On the other hand, photon yield is optimized at high laser intensities when the Rayleigh length is comparable to the pulse length $z_R \sim c\tau$, corresponding to $w_0 \approx 1\mu m$ in the grey area of Fig. 1(a). However, focusing laser to small spot size of $w_0 \approx 1\mu m$ results in higher field strength and broadening of the ICS spectrum. Therefore, the gap between high photon yield and low bandwidth cannot be bridged using ordinary focusing optics, for instance, a parabolic plasma mirror [31–33].

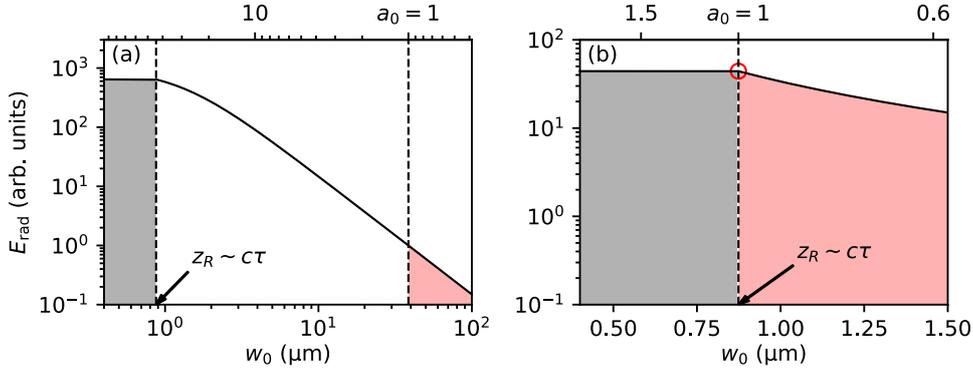

Fig. 1 The schematic of the relation between the radiation yield $E_{\rm rad}$ in ICS and the beam waist of the colliding laser pulse $w_0$ for a 50TW/30fs laser in the case of ordinary optical focusing (a) and spatial-temporal coupling (b), as a function of the beam waist/field strength. The radiation yield is normalized to that of $a_0 = 1$. In either case, yield saturates beginning at $z_R \sim c\tau$ (gray) while the linear ICS corresponds to $a_0 \leq 1$ (red). Optimal gamma radiation is achieved for $a_0 = 1$ and $z_R \sim c\tau$, as shown by the red circle in (b).

We propose a solution to the dilemma using a spatial-temporal coupling geometry. Here a parabolic mirror is azimuthally decomposed into $N$ sections and stretched along the laser axis with a chosen spacing $d$, forming a helical parabolic mirror, as illustrated in Fig. 2(a). When impinging onto the divided parabola, a fraction of the incident laser pulse is reflected and focused to the collision point every period of time. Adjacent sub-parabolas are displaced by a distance of $d$, therefore the time of each sub-pulse arriving at its focus is delayed by $d/c$. One can see in Fig. 2(b-c) that a train of identical sub-pulses are generated with intensity peak shifting backwards along with the electron beam. Each carries the same field amplitude below unity. The proposed geometry is essentially an azimuthal spatial-temporal convertor that decouples the focal field strength from laser spot size such that the light intensity from every focusing element can be programmed to assure linear ICS and the matching condition between the beam waist the electron beam radius. Since each focal point is no longer fixed but moving against the laser propagation direction at the speed of light, it is synchronized with the trailing electron beam (see Fig. 2(c)) to guarantee optimized collision set in Fig. 1(b): the monochromatic requirement $a_0 = 1$, the saturation condition $z_R \sim c\tau$.

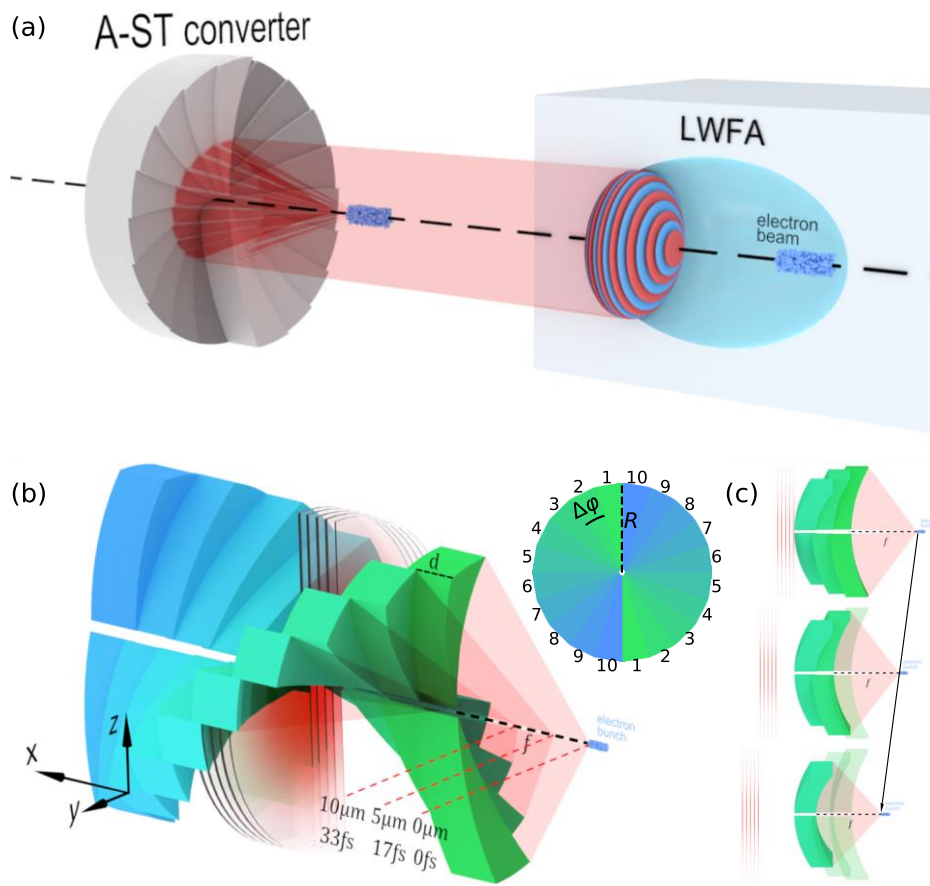

Fig. 2. The azimuthal spatial-temporal coupling geometry. (a) The schematic geometry of the collision of an electron beam from LWFA to a pulse train reflected by the A-ST convertor (b) The mirror structure and sub-pulses during reflection (green for nearer; blue for further). The front-view of the mirror is shown in the upper-right. (c) The synchronized collision between the electron beam and the first 3 sub-pulses with focal position moving inversely to its wave vector at the speed of light. Each mirror component is subdivided into a pair to assure symmetric collision force along the transverse directions.

It has been shown that spatial-temporal coupling can be realized via spatial-dispersive lenses [34,35], axially delayed mirror [36] and axial-parabola [37,38], offering a superluminal laser focus-a promising approach to drive dephasingless wakefield acceleration [36]. All existing spatial-temporal concepts essentially depend on radially delayed focusing, namely, the radial spatial-temporal (R-ST) convertor, which can be simplified to the structure illustrated in Fig. 3(b). The sub-pulses generated by these axially symmetric parabolas are reflected or transformed at different radii, therefore having different divergences and focal depths or Rayleigh lengths, e.g., $z_R^{(3)} > z_R^{(2)} > z_R^{(1)}$ in Fig. 3(b) where $z_R = \pi w_0^2/\lambda$ is the Rayleigh length and $w_0$ is the beam waist. Thus, in either LWFA or ICS only a few sub-pulses work in optimal conditions since the pulse Rayleigh length should match plasma density required by optimal LWFA conditions [39,40] and the pulse divergence angles should remain constant for monochromatic gamma beam in ICS. The majority rest are out of match. Unlike in the R-ST geometry, the proposed A-ST convertor in principle generates identical sub-pulses with the same spot size, field strength and polarization near the axis, as shown in Fig. 3(a), providing identical colliding pulses for the Compton scattering and invariant driving pulse for LWFA and hence ensuring optimized interaction for all sub-pulses.

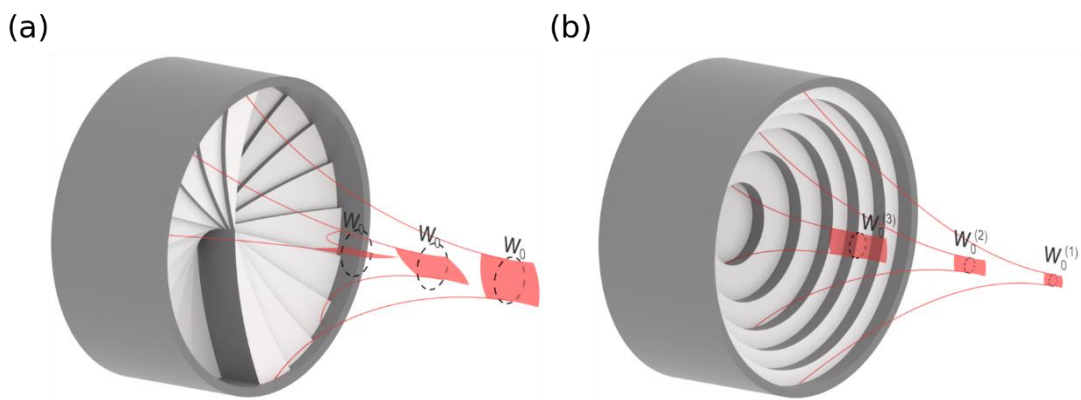

Fig. 3. Azimuthal spatial-temporal convertor generates sub-pulses of identical focal length (a) while radial spatial-temporal convertor produces different focal lengths for each sub-pulse (b).

## II. RESULTS

The idea is demonstrated by considering the ICS process between an 500MeV electron beam and the driving laser pulse of $a_0 = 2$ and wavelength of 800 nm. The electron beam contains 4.4pC total charge distributed in a cylinder of 0.5μm in radius and 2μm in length, corresponding to 1% energy spread and 1 mrad angular divergence [41,42]. Their interaction is mimicked in full three-dimensional particle-in-cell simulations to include the diffraction effects and the laser-plasma interactions at such extreme field. The radiation from Compton scattering is calculated by resolving the Lienard-Wiechert potential of each traced electron (see in Methods). For synchronization of the trailing electron and reflected laser pulse, we constrain the laser spot size to $w_0 = 10\,\mu m$ and the focal length of the reflecting mirror to be $f = 30\,\mu m$. For typical LWFA configuration, such driving laser pulse of $a_0 = 2$ and pulse duration to 27 fs corresponds to peak power of 13.4 TW and beam energy of 358 mJ.

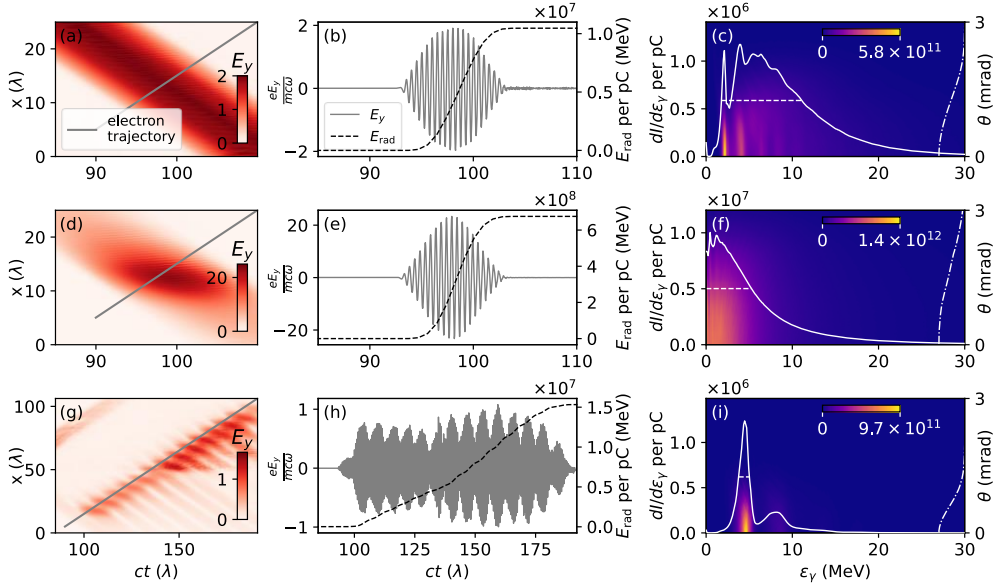

Fig. 4. Field strengths and ICS spectrums. (a) The on-axis field strengths $E_y$ reflected by a flat mirror; (b) the electric fields experienced by an electron on the axis (grey-solid), corresponding to the grey line in (a) and the radiation yield of the electron beam (black-dashed); (c) the produced spectrum $\frac{d^2 I}{d\varepsilon_\gamma d\Omega}$ per pC in units of $rad^{-2}$ where the white-solid lines are the energy spectrums $\frac{dI}{d\varepsilon_\gamma}$ and the white-dashed lines are the FWHM and the peak photon energy. (d-f) The results for a full parabolic mirror. (g-i) The results for A-ST geometry where the first mirror pair is set flat.

We compare the results with the quasi-plane-wave and the normal PM with fixed focus. For the latter, the trailing e-beam is displaced by $l_e = 2f = 60\,\mu m$ with respect to the laser pulse, ensuring matched collision. Note that the distance $l_e$ can be varied by controlling the LWFA conditions like plasma density and laser spot size [40]. The on-axis photon energy from linear ICS with 500 MeV electron is 5.94 MeV. In general, the central beam energy is lower than the on-axis value after integrating over a range of emitting angles. The intense laser field can also induce significant red-shift. In the flat-mirror geometry (see in Fig. 4(a-c)), electrons undergo oscillations at peak field strength of $a_0 = 2$, a typical field strength employed in the single pulse reflection-collision geometry [22]. It produces a broadened and red-shifted spectrum, peaking at 3.88MeV with bandwidth of $\Delta \varepsilon_\gamma = 9.32$ MeV at FWHM (full-width-half-maximum), divergence angle of $\Delta\theta \approx 2.61$ mrad (FWHM). Overall $E_{rad} = 1.05 \times 10^7$ MeV/pC energies are radiated, corresponding to about $N_\gamma = E_{rad}/\langle \varepsilon_\gamma \rangle \approx 1.18 \times 10^6$ photons/pC.

The highest yield is expected by focusing the whole pulse into one spot with a parabolic mirror of $f = 30$ μm (equivalent to A-ST convertor of $N = 1$ mirror pair). In this case, the focal field strength is magnified by an order of 12.5 to $a_0 \approx 25$. Although it produces much more radiated energy $E_{\text{rad}} = 6.76 \times 10^8$ MeV/pC, the strong nonlinear effect and radiation recoil result in extremely red-shifted spectrum, $\varepsilon_{\text{peak}} \approx 0.47$ MeV. Photons are emitted to a much larger divergence angle $\Delta\theta \approx 3.72$ mrad. The energy width is $\Delta\varepsilon_\gamma = 5.41$ MeV, containing abundant low-energy photons as shown in Fig. 4(d)&(f), corresponding to $N_\gamma = 1.05 \times 10^8$ photons/pC when counting gamma photons greater than 1 MeV.

When switching to the A-ST geometry, the proposed azimuthal structure generates sub-pulses of similar duration and field strength which are fully synchronized with the collision process (see in Fig. 4(g)). Fourteen sub-pulses are aligned in Fig. 4(h) carrying maximum field amplitude of $a_0 \approx 1$, where $N = 15$ mirror pairs are utilized and the first section of mirror pair is flattened to remove the first pulse that is beyond $a_0 = 1$, which will be discussed in Section III B. Consequently, electrons experience about 14 times longer period of oscillation under the same field strengths as compared to the quasi-plane-wave case. The radiation yield is boosted to $E_{\text{rad}} = 1.53 \times 10^7$ MeV/pC, corresponding to $N_\gamma = 3.43 \times 10^6$ photons/pC peaking at 4.47 MeV. The bandwidth is suppressed to $\Delta\varepsilon_\gamma = 1.31$ MeV, which is 7-fold suppression of the flat mirror geometry. The photon beam is much more collimated to $\Delta\theta \approx 1.35$ mrad, as shown in Fig. 4(i). In terms of brilliance, measured by $N_\gamma/(\Delta\varepsilon_\gamma \Delta\theta^2)$, the A-ST convertor reaches $1.08 \times 10^6$ photons/pC/MeV/mrad2, almost 60 times of the flat mirror geometry that produces $1.86 \times 10^4$ photons/pC/MeV/mrad2.

On the other hand, considering the laser efficiency, the A-ST convertor produces $9.58 \times 10^6$ photons/pC/J at 4.47 MeV out of an 358 mJ laser pulse, which is 21 times of $4.5 \times 10^5$ photons/pC/J at up to 1MeV out of a laser pulse of 0.13 J and $a_0 = 0.3$ [43], 63 times of $1.5 \times 10^5$ photons/pC/J at 15keV out of a laser pulse of 300mJ and $a_0 = 0.85$ [44] and 17 times of $5.05 \times 10^5$ photons/pC/J at up to 2 MeV out of a 3.3J laser pulse of $a_0 = 1.3$ [23].

Comparing the three cases, it is obvious that the A-ST geometry realizes linear ICS using full laser power, which is otherwise prohibited by single-focus optics. It successfully boosts the radiation yield and retains the quasi-mono-chromatic feature. On top of that the beam collimation is also greatly improved.

### III. DISCUSSION AND CONCLUSION

#### A. Scaling law of the mirror pairs

The photon spectrums of different pair numbers $N$ are presented in Fig. 5(a). The peak photon energy increases for higher $N$ and the bandwidth is narrowed simultaneously. The spectrum exhibits quasi-mono-chromatic feature starting from $N = 10$ and the energy spread is further suppressed by flattening the first mirror pair (without first pair), which will be explained in the next section. From Fig. 5(b)&(c), we find that the brilliance of the A-ST convertor peaks at $N = 15$ where the spectrum bandwidth and beam collimation are also optimized. Here high energy photon radiation is hindered for the undivided parabolic mirror ($N = 1$) case, as the electron energy is significantly depleted from radiation-reaction. These lead to relatively small bandwidth but at the cost of strong red-shift and loss of high-energy photons. The radiation yield is also dependent on the focal length. One sees highest radiation energy around $f \approx 27\ \mu m$, shown in Fig. 5(d), a slight shift from the designed $f = 30\mu m$. Because in simulations we use a Gaussian pulse of $w_0 = 10\ \mu m$, which is not an ideal plane wave. As for the central photon energy and bandwidth, both remain stable for different focal lengths, as shown in Fig. 5(d-e). We notice that the trend in Fig. 5(d) indicates a focal window of $\approx 6\mu m$ at FWHM. It defines the tolerance on focusing fluctuation in the current set-up.

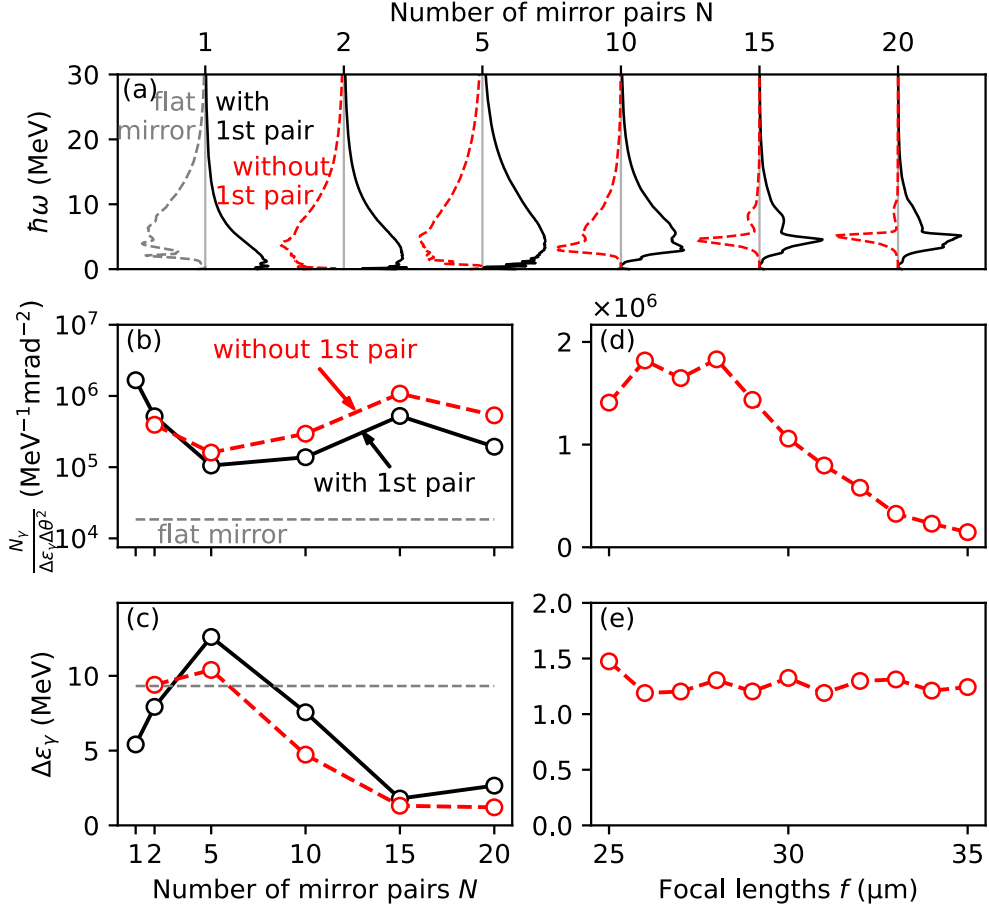

Fig. 5. The scaling law for the number of parabolic mirror pairs. (a) Spectrums for different pair numbers. The black lines denote A-ST convertors with the first mirror pairs; the red dashed lines denote A-ST convertors with the first pairs flattened; the grey dashed line is the result of the flat mirror. (b) The radiation brilliance measured by $\frac{dN_\gamma}{\Delta\varepsilon_\gamma \Delta\theta^2}$ per pC, (c) the bandwidth $\Delta\varepsilon_\gamma$. (d-e) The results for different focal lengths of $N = 15$ corresponding to (b-c). The solid-black (dashed-red) lines are the results with (without) the first mirror pairs, and the dashed-grey line corresponds to the flat mirror.

## B. Diffraction loss from the mirror pairs

We further discuss the diffraction effect present in the A-ST geometry. The wavefronts of the incident laser pulse during interaction with the A-ST convertor and the on-axis field strengths are shown in Fig. 6(a-d) for $N = 15$ and Fig. 6(e-h) for $N = 5$. The first mirror pair experience an undisturbed wavefront, after which the wavefront is then disturbed by diffraction effect from the finite size of the mirror pairs and the pulses are relatively weaker than the first pulse as shown by the black lines in Fig. 6(d)&(h). The diffraction is stronger for smaller mirror pairs, as indicated by the difference between the amplitudes. We then flatten the first mirror pair to eliminate the nonlinear effect induced by the first sub-pulse. Such micro structure can be fabricated via 3D printing techniques [45] for experimental consideration.

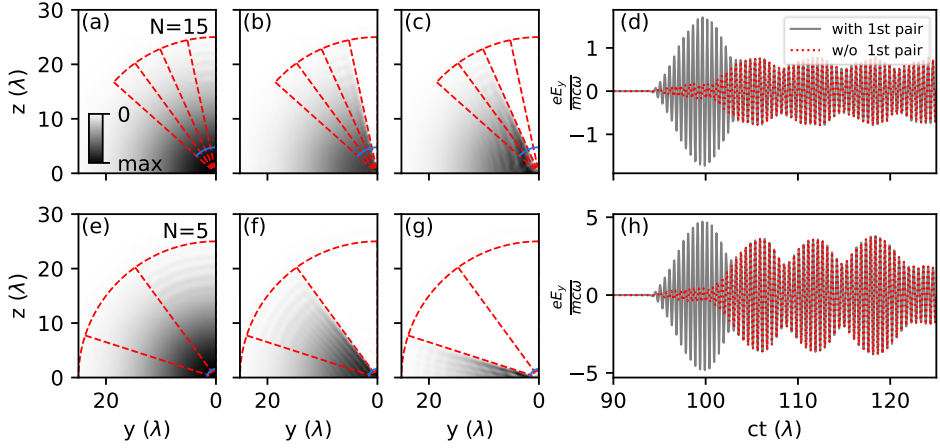

Fig. 6. Wavefronts of the incident laser and on-axis field strengths. (a-c) The field strengths in the plane of phase 0 and (d) on-axis field strengths for (a-d) $N = 15$ mirror pairs and (e-h) $N = 5$ mirror pairs. The dashed-red lines in (a-c) and (e-h) are the outlines of the mirrors and the blue arcs indicate the radius when the arc length of a mirror equals the laser wavelength, which measures the region of strong diffraction effect. The dotted-red (solid-black) lines in (d)&(h) are the on-axis field strengths experienced by an electron with (without) the first mirror pair flattened.

### C. Diffraction loss during focusing

Taking into account the diffraction effect, the field strengths of the sub-pulses for different pair numbers are shown in Fig. 7. The f-number of a parabolic mirror is $f_\# = \frac{f}{2R}$ with $R$ the mirror radius. For an undivided parabolic mirror, the focal spot size should be $\sim \lambda f_\#$. For parabolic mirror pair, the radial size remains $2R$ but the azimuthal size decreases to $\sim \frac{2R}{N}$, which expands the focal spot size in the azimuthal direction to $\sim \lambda \frac{f}{2R/N} = \lambda N f_\#$. Since each sub-pulse shares $1/N$ of the laser power $P \sim a_0^2 w_0^2 \sim 1/N$ and $w_0$ scales as $w_0^2 \sim \lambda f_\# \cdot \lambda N f_\# \sim N$, the focal field strength therefore scales as $1/N$. In other words, the finite size of the mirror pairs and consequent diffraction produce a scaling $\sim 1/N$ (dashed-black line in Fig. 7) rather than $1/\sqrt{N}$. From another point of view, since the Maxwell equation is linear, the focal field of an undivided parabolic mirror can be linearly decomposed into multiple combination of fields from the subdivisions of the mirror, which are the mirror pairs of the A-ST convertor. Therefore, the focal field strengths of the sub-pulses can be predicted by

$$a_0^{foc}(N) = \frac{a_0^{foc}(1)}{N} \quad (1)$$

which is exactly the black-dashed line in Fig. 7. The above relationship provides a simple rule to design the structure of A-ST convertor for high quality gamma-ray beam generation by requiring $N \sim a_0^{foc}(1)$.

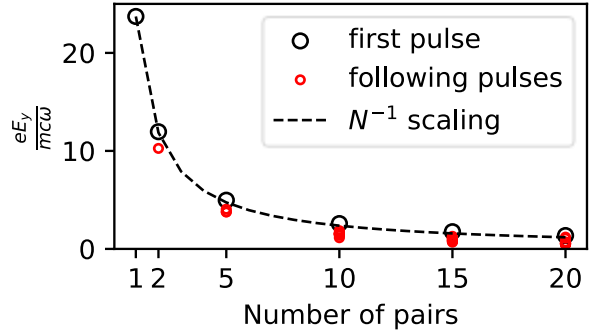

Fig. 7. The on-axis field strengths of the pulse train for different number of mirror pairs. The black circle denotes the field strength of the first sub-pulse and the smaller red circles denote the following pulses. The dashed line is the field strengths fitted with $1/N$.

### D. General applications

For more general purposes, the A-ST convertor can be up-scaled to larger sizes to interact with laser pulses below the damage threshold of the mirror, which is also easier to fabricate. In this geometry, an independent pulse can be used in the collision with the electron bunch, relaxing the constrains on the laser pulse from the output of the LWFA process. Besides, this geometry re-distributes the laser energy and generates co-moving focuses behind the laser pulse, which satisfies the acquirements of applications like dephasingless LWFA [36] and ponderomotive-controlled radiation [46]. For micro-meter-scale A-ST convertors, the microstructure can be manufactured with the state-of-the-art 3D-print techniques [45], which can be collimated by detecting signals passing through the passage for the electron beam (see Fig. 2(b)).

In conclusion, the proposed A-ST geometry can transform the incident laser pulse into a train of almost identical pulses, which controls the field strength below the nonlinear threshold of inverse Compton scattering process without compromising the radiation yield in the all-optical Compton source. It significantly suppresses the bandwidth of the generated radiation spectrum and achieves monochromatic output.

## IV. METHODS

### A. Estimation of radiation yield

The radiation yield in Fig. 1 is estimated by integrating the RR force in the collision between an electron of $\gamma_0 = 1000$ and a laser pulse of 30fs at FWHM. For highly relativistic electrons ($\gamma \gg 1$), the RR force, taken from the Landau-Lifshitz equation [14], is approximated by

$$\mathbf{F}_{RR} \approx -\frac{2}{3}\frac{e^4}{m^2c^5}\gamma^2[(\mathbf{E} + \mathbf{v} \times \mathbf{B})^2 - (\boldsymbol{\beta} \cdot \mathbf{E})^2]\mathbf{v}, \qquad (2)$$

where $\boldsymbol{\beta} = \mathbf{v}/c$ and $\mathbf{v}$ is the electron velocity. Since $\gamma \gg a_0$ is considered, the electron motion in the laser field is simplified to a 1-D trajectory and $\mathbf{F}_{RR} \approx -\frac{8}{3}\frac{e^2\omega^2}{c^2}\gamma^2 a_0^2 \mathbf{v}$. The radiation power becomes $\frac{d\gamma mc^2}{dt} = \mathbf{F}_{RR} \cdot \mathbf{v} \approx -\frac{8}{3}\frac{e^2\omega^2}{c}\gamma^2 a_0^2(t)$ where $a_0(t)$ is the field strength experienced by the electron, which is then integrated to $\int_{\gamma_0}^{\gamma_f} \gamma^{-2} d\gamma = -\frac{8}{3}\frac{e^2\omega^2}{mc^3}\int_{t_i}^{t_f} a_0^2(t) dt = -I(t_i, t_f)$, and finally we have $\gamma_f = [\gamma_0^{-1} + I(t_i, t_f)]^{-1}$. The radiation yield is estimated by $(\gamma_0 - \gamma_f) mc^2$. The integration limits are determined by the interaction period that decreases to $2\frac{Z_R}{c}$ when the Rayleigh length is smaller than the pulse length for extremely tight focusing, which results in the saturation plateau in Fig. 1.

### B. Simulation set-up

The simulation is carried out via the Smilei code [47] with cell-size of $0.05\lambda \times 0.17\lambda \times 0.17\lambda$ in the x, y and z dimensions with laser pulse in the x-direction. The size of the simulation box is adjusted for different number of mirror pairs so long as the mirror plus its focal length can fit into the simulation domain. For example, the box size is $117.6\mu m \times 44\mu m \times 44\mu m$ for $N = 15$ mirror pairs, corresponding to $2940 \times 320 \times 320$ grids.

The A-ST convertor with focal length of $f = 30\mu m$ is set to be pre-ionized in the simulation and its electron number density is $100 n_c$ with 16 macro-particles per cell. Here $n_c = \varepsilon_0 m\omega^2/e^2$ is the critical plasma density with $\varepsilon_0$ being the vacuum permittivity. The convertor is generated by subdividing a parabolic mirror into $2N$ sections, or $N$ opposite pairs, and displacing adjacent mirror pairs by $d = 5\mu m$. A tunnel of radius of $1\mu m$ through the mirror is reserved for electron propagation.

The electron beam enters the simulation box from the boundary along the x-axis following the laser pulse. The pulse profile is defined by $a_0 \sin(\pi t/2\tau) e^{-r^2/w_0^2}$ where $a_0 = 2$, $w_0 = 10\mu m$ and $\tau = 27fs$. The spacing between the beam and the peak of laser profile is $l_e = 2f$ required by the time delay introduced by parabolic mirror. The central energy of the electron beam is 500 MeV with 1% energy spread and angular divergence of 1 mrad at $e^{-1}$ of a Gaussian profile. The electron beam is uniformly distributed in a cylinder of 0.5μm in radius and 2μm in length, which represents 4.4pC (4.65pC in the simulation due to grid effect) of electrons that corresponds to the number density of $1.744 \times 10^{25}$ m$^{-3}$ with 10 macro-particles per cell. Take LWFA for example, the spacing $l_e$ between the laser peak and the beam center is dependent on the size of the bubble $r_b$ [40] as the electron beam is accelerated at the tail of the bubble. This spacing can be tuned by changing the LWFA conditions, e.g., the plasma density $n_0$, the field strength $a_0$, and the beam waist $w_0$ of the driving laser.

## C. Radiation spectrum

The radiation fields are obtained by calculating the Lienard-Wiechert potential [15] in specific directions following the electron trajectories $(\mathbf{x}, \mathbf{p})$ from PIC simulations:

$$\mathbf{E} \approx \frac{e}{c^2 R} \frac{\mathbf{n} \times \left[\left(\mathbf{n} - \frac{\mathbf{v}}{c}\right) \times \dot{\mathbf{v}}\right]}{\left(1 - \mathbf{n} \cdot \frac{\mathbf{v}}{c}\right)^3}, \tag{3}$$

where $R$ is the distance from the electron to the observation point, $\mathbf{n}$ is the direction to the observation point, $\mathbf{v}$ the velocity of electron and $\dot{\mathbf{v}}$ the acceleration. The right-hand side is evaluated at time steps of electron and the left-hand side is defined at time steps of $t' = t + R/c$. Then we calculate the spectrum via Fourier transformation of the electric fields in the time domain. Since $t'$ depends on $R(t)$ and is not uniformly distributed, we first interpolate the radiation fields on a uniform time grid, then we perform fast Fourier transformation to the interpolated radiation fields. One can refer to the thesis [48] for numerical details.

Since the pulse length of the radiation fields are compressed by a factor of $4\gamma^2$ compared to the laser pulse length, i.e., $\frac{c\tau}{4\gamma^2} \sim 2\text{pm}$, much smaller than the average distance between electrons, i.e., $n_e^{-1/3} \approx 4\text{nm}$, the radiation is highly incoherent for highly relativistic electrons and the spectrum of each trajectory can be simply summed up without considering coherent effect.

The electron trajectories taken from PIC simulations inherently include the radiation-reaction effect in Eq. (2). This is necessary for the case of $a_0 = 25$. For electron energy of $\gamma = O(10^3)$ and field strength of $a = O(1)$, we have the quantum parameter [49] $\chi = O(10^{-2})$ that indicates the nonlinear quantum effects are weak [50–52]. Therefore classical calculations are sufficient to deal with the radiation spectrum. In the case of $a_0 \approx 25$ and $\chi_e \approx 0.15$, where quantum effects may arise, we find the classical estimation does not deviate much from the QED Monte-Carlo calculation of the Smilei code. For consistency consideration, we keep the classical calculations in our results.

## ACKNOWLEDGMENTS

We acknowledge funding from National Science Foundation of China (Nos.11875307,11935008,11804348); Strategic Priority Research Program of Chinese Academy of Sciences grant (No. XDB16010000).

[1] C. R. Howell, M. W. Ahmed, A. Afanasev, D. Alesini, J. R. M. Annand, A. Aprahamian, D. L. Balabanski, S. V. Benson, A. Bernstein, C. R. Brune, J. Byrd, B. E. Carlsten, A. E. Champagne, S. Chattopadhyay, D. Davis, E. J. Downie, J. M. Durham, G. Feldman, H. Gao, C. G. R. Geddes, H. W. Grießhammer, R. Hajima, H. Hao, D. Hornidge, J. Isaak, R. V. F. Janssens, D. P. Kendellen, M. A. Kovash, P. P. Martel, U.-G. Meißner, R. Miskimen, B. Pasquini, D. R. Phillips, N. Pietralla, D. Savran, M. R. Schindler, M. H. Sikora, W. M. Snow, R. P. Springer, C. Sun, C. Tang, B. Tiburzi, A. P. Tonchev, W. Tornow, C. A. Ur, D. Wang, H. R. Weller, V. Werner, Y. K. Wu, J. Yan, Z. Zhao, A. Zilges, and F. Zomer, *International Workshop on next Generation Gamma-Ray Source*, J. Phys. G: Nucl. Part. Phys. **49**, 010502 (2021).

[2] F. R. Elder, R. V. Langmuir, and H. C. Pollock, *Radiation from Electrons Accelerated in a Synchrotron*, Phys. Rev. **74**, 52 (1948).
[3] I. M. Ternov, *Synchrotron Radiation*, Phys.-Usp. **38**, 409 (1995).
[4] S. S. R. Lightsource (SSRL), *As a National User Facility the Stanford Synchrotron Radiation Lightsource Invites Scientists from All over the World to Use Our Intense X-Ray Beam and World-Class Instruments for Their Experiments.*, http://www-ssrl.slac.stanford.edu/content.
[5] *DESY PHOTON SCIENCE*, https://photon-science.desy.de/.
[6] *HZDR – Helmholtz-Zentrum Dresden-Rossendorf*, https://www.hzdr.de.
[7] J. M. J. Madey, *Stimulated Emission of Bremsstrahlung in a Periodic Magnetic Field*, Journal of Applied Physics **42**, 1906 (1971).
[8] *FLASH*, https://photon-science.desy.de/facilities/flash/index_eng.html.
[9] *European XFEL*, https://www.xfel.eu.
[10] T. ERBER, *High-Energy Electromagnetic Conversion Processes in Intense Magnetic Fields*, Rev. Mod. Phys. **38**, 626 (1966).
[11] S. Li, B. Shen, J. Xu, T. Xu, Y. Yu, J. Li, X. Lu, C. Wang, X. Wang, X. Liang, Y. Leng, R. Li, and Z. Xu, *Ultrafast Multi-MeV Gamma-Ray Beam Produced by Laser-Accelerated Electrons*, Physics of Plasmas **24**, 093104 (2017).
[12] R. H. Milburn, *Electron Scattering by an Intense Polarized Photon Field*, Phys. Rev. Lett. **10**, 75 (1963).
[13] F. R. Arutyunian and V. A. Tumanian, *The Compton Effect on Relativistic Electrons and the Possibility of Obtaining High Energy Beams*, Physics Letters **4**, 176 (1963).
[14] L. D. Landau and E. M. Lifshitz, *The Classical Theory of Fields* (Pergamon Press, Oxford, 1971).
[15] J. D. Jackson, *Classical Electrodynamics 3rd Edition* (Wiley, New York, 1998).
[16] P. Sprangle, A. Ting, E. Esarey, and A. Fisher, *Tunable, Short Pulse Hard X-rays from a Compact Laser Synchrotron Source*, Journal of Applied Physics **72**, 5032 (1992).
[17] K.-J. Kim, S. Chattopadhyay, and C. V. Shank, *Generation of Femtosecond X-Rays by 90° Thomson Scattering*, Nuclear Instruments and Methods in Physics Research Section A: Accelerators, Spectrometers, Detectors and Associated Equipment **341**, 351 (1994).
[18] J. Gao, *Thomson Scattering from Ultrashort and Ultraintense Laser Pulses*, Phys. Rev. Lett. **93**, 243001 (2004).
[19] G. A. Krafft, A. Doyuran, and J. B. Rosenzweig, *Pulsed-Laser Nonlinear Thomson Scattering for General Scattering Geometries*, Phys. Rev. E **72**, 056502 (2005).
[20] P. Catravas, E. Esarey, and W. P. Leemans, *Femtosecond X-Rays from Thomson Scattering Using Laser Wakefield Accelerators*, Meas. Sci. Technol. **12**, 1828 (2001).
[21] F. V. Hartemann, D. J. Gibson, W. J. Brown, A. Rousse, K. T. Phuoc, V. Mallka, J. Faure, and A. Pukhov, *Compton Scattering X-Ray Sources Driven by Laser Wakefield Acceleration*, Phys. Rev. ST Accel. Beams **10**, 011301 (2007).
[22] K. Ta Phuoc, S. Corde, C. Thaury, V. Malka, A. Tafzi, J. P. Goddet, R. C. Shah, S. Sebban, and A. Rousse, *All-Optical Compton Gamma-Ray Source*, Nature Photonics **6**, 5 (2012).
[23] C. Yu, R. Qi, W. Wang, J. Liu, W. Li, C. Wang, Z. Zhang, J. Liu, Z. Qin, M. Fang, K. Feng, Y. Wu, Y. Tian, Y. Xu, F. Wu, Y. Leng, X. Weng, J. Wang, F. Wei, Y. Yi, Z. Song, R. Li, and Z. Xu, *Ultrahigh Brilliance Quasi-Monochromatic MeV*

γ-*Rays Based on Self-Synchronized All-Optical Compton Scattering*, Scientific Reports **6**, 1 (2016).
[24] H.-E. Tsai, X. Wang, J. M. Shaw, Z. Li, A. V. Arefiev, X. Zhang, R. Zgadzaj, W. Henderson, V. Khudik, G. Shvets, and M. C. Downer, *Compact Tunable Compton X-Ray Source from Laser-Plasma Accelerator and Plasma Mirror*, Physics of Plasmas **22**, 023106 (2015).
[25] D. P. Umstadter, *All-Laser-Driven Thomson X-Ray Sources*, Contemporary Physics **56**, 417 (2015).
[26] A. Döpp, E. Guillaume, C. Thaury, J. Gautier, I. Andriyash, A. Lifschitz, V. Malka, A. Rousse, and K. T. Phuoc, *An All-Optical Compton Source for Single-Exposure x-Ray Imaging*, Plasma Phys. Control. Fusion **58**, 034005 (2016).
[27] H. R. Weller, M. W. Ahmed, H. Gao, W. Tornow, Y. K. Wu, M. Gai, and R. Miskimen, *Research Opportunities at the Upgraded HIγS Facility*, Progress in Particle and Nuclear Physics **62**, 257 (2009).
[28] Y. Sakai, I. Pogorelsky, O. Williams, F. O'Shea, S. Barber, I. Gadjev, J. Duris, P. Musumeci, M. Fedurin, A. Korostyshevsky, B. Malone, C. Swinson, G. Stenby, K. Kusche, M. Babzien, M. Montemagno, P. Jacob, Z. Zhong, M. Polyanskiy, V. Yakimenko, and J. Rosenzweig, *Observation of Redshifting and Harmonic Radiation in Inverse Compton Scattering*, Phys. Rev. Accel. Beams **18**, 060702 (2015).
[29] G. Sarri, D. J. Corvan, W. Schumaker, J. M. Cole, A. Di Piazza, H. Ahmed, C. Harvey, C. H. Keitel, K. Krushelnick, S. P. D. Mangles, Z. Najmudin, D. Symes, A. G. R. Thomas, M. Yeung, Z. Zhao, and M. Zepf, *Ultrahigh Brilliance Multi-MeV γ-Ray Beams from Nonlinear Relativistic Thomson Scattering*, Phys. Rev. Lett. **113**, 224801 (2014).
[30] W. Yan, C. Fruhling, G. Golovin, D. Haden, J. Luo, P. Zhang, B. Zhao, J. Zhang, C. Liu, M. Chen, S. Chen, S. Banerjee, and D. Umstadter, *High-Order Multiphoton Thomson Scattering*, Nature Photonics **11**, 8 (2017).
[31] S. V. Bulanov, T. Esirkepov, and T. Tajima, *Light Intensification towards the Schwinger Limit*, Phys. Rev. Lett. **91**, 085001 (2003).
[32] H. Vincenti, *Achieving Extreme Light Intensities Using Optically Curved Relativistic Plasma Mirrors*, Phys. Rev. Lett. **123**, 105001 (2019).
[33] F. Quéré and H. Vincenti, *Reflecting Petawatt Lasers off Relativistic Plasma Mirrors: A Realistic Path to the Schwinger Limit*, High Power Laser Science and Engineering **9**, (2021).
[34] D. H. Froula, D. Turnbull, A. S. Davies, T. J. Kessler, D. Haberberger, J. P. Palastro, S.-W. Bahk, I. A. Begishev, R. Boni, S. Bucht, J. Katz, and J. L. Shaw, *Spatiotemporal Control of Laser Intensity*, Nature Photonics **12**, 5 (2018).
[35] D. H. Froula, J. P. Palastro, D. Turnbull, A. Davies, L. Nguyen, A. Howard, D. Ramsey, P. Franke, S.-W. Bahk, I. A. Begishev, R. Boni, J. Bromage, S. Bucht, R. K. Follett, D. Haberberger, G. W. Jenkins, J. Katz, T. J. Kessler, J. L. Shaw, and J. Vieira, *Flying Focus: Spatial and Temporal Control of Intensity for Laser-Based Applications*, Physics of Plasmas **26**, 032109 (2019).
[36] J. P. Palastro, J. L. Shaw, P. Franke, D. Ramsey, T. T. Simpson, and D. H. Froula, *Dephasingless Laser Wakefield Acceleration*, Phys. Rev. Lett. **124**, 134802 (2020).
[37] K. Oubrerie, I. A. Andriyash, S. Smartsev, V. Malka, and C. Thaury, *Axiparabola: A New Tool for High-Intensity Optics*, ArXiv:2106.06802 [Physics] (2021).
[38] S. Smartsev, C. Caizergues, K. Oubrerie, J. Gautier, J.-P. Goddet, A. Tafzi, K. T. Phuoc, V. Malka, and C. Thaury, *Axiparabola: A Long-Focal-Depth, High-Resolution Mirror for Broadband High-Intensity Lasers*, Opt. Lett., OL **44**, 3414

(2019).

[39] C. G. Durfee and H. M. Milchberg, *Light Pipe for High Intensity Laser Pulses*, Phys. Rev. Lett. **71**, 2409 (1993).

[40] W. Lu, C. Huang, M. Zhou, M. Tzoufras, F. S. Tsung, W. B. Mori, and T. Katsouleas, *A Nonlinear Theory for Multidimensional Relativistic Plasma Wave Wakefields*, Physics of Plasmas **13**, 056709 (2006).

[41] T. Tajima and J. M. Dawson, *Laser Electron Accelerator*, Phys. Rev. Lett. **43**, 267 (1979).

[42] A. J. Gonsalves, K. Nakamura, J. Daniels, C. Benedetti, C. Pieronek, T. C. H. de Raadt, S. Steinke, J. H. Bin, S. S. Bulanov, J. van Tilborg, C. G. R. Geddes, C. B. Schroeder, Cs. Tóth, E. Esarey, K. Swanson, L. Fan-Chiang, G. Bagdasarov, N. Bobrova, V. Gasilov, G. Korn, P. Sasorov, and W. P. Leemans, *Petawatt Laser Guiding and Electron Beam Acceleration to 8 GeV in a Laser-Heated Capillary Discharge Waveguide*, Phys. Rev. Lett. **122**, 084801 (2019).

[43] N. D. Powers, I. Ghebregziabher, G. Golovin, C. Liu, S. Chen, S. Banerjee, J. Zhang, and D. P. Umstadter, *Quasi-Monoenergetic and Tunable X-Rays from a Laser-Driven Compton Light Source*, Nature Photonics **8**, 1 (2014).

[44] K. Khrennikov, J. Wenz, A. Buck, J. Xu, M. Heigoldt, L. Veisz, and S. Karsch, *Tunable All-Optical Quasimonochromatic Thomson X-Ray Source in the Nonlinear Regime*, Phys. Rev. Lett. **114**, 195003 (2015).

[45] H. Gao, Y. Hu, Y. Xuan, J. Li, Y. Yang, R. V. Martinez, C. Li, J. Luo, M. Qi, and G. J. Cheng, *Large-Scale Nanoshaping of Ultrasmooth 3D Crystalline Metallic Structures*, Science **346**, 1352 (2014).

[46] D. Ramsey, B. Malaca, A. Di Piazza, M. F. P. Franke, D. H. Froula, M. Pardal, T. T. Simpson, J. Vieira, K. Weichman, and J. P. Palastro, *Nonlinear Thomson Scattering with Ponderomotive Control*, ArXiv:2108.04044 [Physics] (2021).

[47] J. Derouillat, A. Beck, F. Pérez, T. Vinci, M. Chiaramello, A. Grassi, M. Flé, G. Bouchard, I. Plotnikov, N. Aunai, J. Dargent, C. Riconda, and M. Grech, *Smilei : A Collaborative, Open-Source, Multi-Purpose Particle-in-Cell Code for Plasma Simulation*, Comput. Phys. Comm. **222**, 351 (2018).

[48] P. Richard, Electromagnetic Radiation from Relativistic Electrons as Characteristic Signature of Their Dynamics, Technische Universität Dresden, 2012.

[49] V. I. Ritus, *Quantum Effects of the Interaction of Elementary Particles with an Intense Electromagnetic Field*, J Russ Laser Res **6**, 497 (1985).

[50] N. Neitz and A. Di Piazza, *Stochasticity Effects in Quantum Radiation Reaction*, Phys. Rev. Lett. **111**, 054802 (2013).

[51] C. N. Harvey, A. Gonoskov, A. Ilderton, and M. Marklund, *Quantum Quenching of Radiation Losses in Short Laser Pulses*, Phys. Rev. Lett. **118**, 105004 (2017).

[52] X. S. Geng, L. L. Ji, B. F. Shen, B. Feng, Z. Guo, Q. Yu, L. G. Zhang, and Z. Z. Xu, *Quantum Reflection above the Classical Radiation-Reaction Barrier in the Quantum Electro-Dynamics Regime*, Commun Phys **2**, 66 (2019).